\documentclass[conference]{IEEEtran}
\ifCLASSINFOpdf
\else
\fi
\usepackage{cite}
\usepackage{xcolor}
\usepackage{graphicx}
\usepackage{amsmath}
\usepackage{amsfonts}
\usepackage{physics}
\usepackage{algorithm}
\usepackage[noend]{algpseudocode}
\usepackage{tabularx}
\usepackage{hhline}
\usepackage[hidelinks]{hyperref}
\usepackage[T1]{fontenc}
\usepackage[utf8]{inputenc}
\usepackage{authblk}
\usepackage{caption}
\usepackage{subcaption}
\makeatletter
\def\BState{\State\hskip-\ALG@thistlm}
\makeatother

\makeatletter
\newsavebox\myboxA
\newsavebox\myboxB
\newlength\mylenA

\newcommand*\xoverline[2][0.75]{%
    \sbox{\myboxA}{$\m@th#2$}%
    \setbox\myboxB\null
    \ht\myboxB=\ht\myboxA%
    \dp\myboxB=\dp\myboxA%
    \wd\myboxB=#1\wd\myboxA
    \sbox\myboxB{$\m@th\overline{\copy\myboxB}$}
    \setlength\mylenA{\the\wd\myboxA}
    \addtolength\mylenA{-\the\wd\myboxB}%
    \ifdim\wd\myboxB<\wd\myboxA%
       \rlap{\hskip 0.5\mylenA\usebox\myboxB}{\usebox\myboxA}%
    \else
        \hskip -0.5\mylenA\rlap{\usebox\myboxA}{\hskip 0.5\mylenA\usebox\myboxB}%
    \fi}
\makeatother

\DeclareMathOperator*{\argmax}{arg\,max}

\IEEEoverridecommandlockouts

\begin{document}
%
\title{An Edge Alignment-based Orientation Selection Method for Neutron Tomography}


\author[1]{Diyu Yang\textsuperscript{\textsection}}
\author[2]{Shimin Tang\textsuperscript{\textsection}}
\author[2]{Singanallur V. Venkatakrishnan}
\author[1]{Mohammad S. N. Chowdhury}
\author[2]{Yuxuan Zhang}
\author[2]{\\Hassina Z. Bilheux}
\author[1]{Gregery T. Buzzard}
\author[1]{Charles A. Bouman}
\affil[1]{Purdue University, West Lafayette, IN  47906}
\affil[2]{Oak Ridge National Laboratory, One Bethel Valley Road, Oak Ridge, TN 37831}

\renewcommand\Authands{ and }

%


\maketitle
\begingroup\renewcommand\thefootnote{\textsection}
\footnotetext{These authors contributed equally.

This manuscript has been authored by UT-Battelle, LLC, under contract DE-AC05-00OR22725 with the US Department of Energy (DOE). The US government retains and the publisher, by accepting the article for publication, acknowledges that the US government retains a nonexclusive, paid-up, irrevocable, worldwide license to publish or reproduce the published form of this manuscript, or allow others to do so, for US government purposes. DOE will provide public access to these results of federally sponsored research in accordance with the DOE Public Access Plan (http://energy.gov/downloads/doe-public-access-plan).}
\endgroup
\begin{abstract}
Neutron computed tomography (nCT) is a 3D characterization technique used to image the internal morphology or chemical composition of samples in biology and materials sciences. 
A typical workflow involves placing the sample in the path of a neutron beam,
acquiring projection data at a predefined set of orientations, and processing the resulting data using an analytic reconstruction algorithm. 
Typical nCT scans require hours to days to complete and are then processed using conventional filtered back-projection (FBP), which performs poorly with sparse views or noisy data.
Hence, the main methods in order to reduce overall acquisition time are the use of an improved sampling strategy combined with the use of advanced reconstruction methods such as model-based iterative reconstruction (MBIR).
In this paper, we propose an adaptive orientation selection method in which an MBIR reconstruction on previously-acquired measurements is used to define an objective function on orientations that balances a data-fitting term promoting edge alignment and a regularization term promoting orientation diversity. 
Using simulated and experimental data, we demonstrate that our method
produces high-quality reconstructions using significantly fewer total measurements than the conventional approach.

\end{abstract}


%
\IEEEpeerreviewmaketitle

\section{Introduction}

Neutron computed tomography (nCT) is a 3D imaging technique used for the non-destructive characterization of samples in biology \cite{bilheux2021neutron,rudolph2021neutron} and materials science \cite{griesche2014three,richards1982neutron}. 
nCT proceeds by placing a sample in the path of a neutron beam, usually at a nuclear reactor or pulsed neutron source facility, and making measurements of the transmitted projection images (also known as views) at various sample orientations. 
Typical nCT systems use a fixed angular step size to rotate the sample about a single axis; the resulting measurements are processed using the filtered back-projection (FBP) algorithm \cite{KakSlaney}. 

The conventional approach for nCT can be extremely time-consuming due to the time required for each view and the number of views required.
Using conventional scanning and reconstruction methods (such as FBP), each view can require from a few minutes (at a reactor) to several hours (at a pulsed source) to obtain sufficient SNR, especially as is the case for hyperspectral systems \cite{woracek20143d,gregg2018tomographic}. 
These factors, plus high operating costs and limited beam time at neutron facilities, create a need for more efficient acquisition and reconstruction methods to shorten nCT scan times.

To reduce the measurement time for nCT, model-based iterative reconstruction (MBIR) methods have been developed and applied to sparse-view interlaced scanning \cite{timbir} or golden ratio \cite{kohler2004projection,kaestner2011spatiotemporal} acquisition protocols. This approach can produce high-quality reconstructions throughout the course of the scan, even before all measurements have been obtained, and can dramatically reduce overall scan time \cite{venkatakrishnan2021improved}.
However, this method is still inefficient because the sampling pattern is fixed and does not take into account the geometry and morphology of the samples that are being measured.



In the context of some computational imaging systems, sample adaptive strategies in which the next measurement is chosen based on previously measured data have been recently proposed. 
Such methods can use either supervised learning based on training data \cite{slads} or unsupervised methods based on a model for the underlying objects to be measured \cite{ziatdinov2022bayesian,seeger2010optimization,haldar2019oedipus}. 
All of these approaches have been developed for scanning microscopy or MRI systems, in which the time to measure is significantly longer than the time to reorient and process the measurements, but none of them are tuned for CT applications.
In contrast, existing dynamic sampling approaches for CT \cite{batenburg2013dynamic,dabravolski2014dynamic} select new samples using information from the existing projection data rather than from the image domain, which limits the information available for sample selection. 

In this paper, we propose an adaptive orientation selection method for nCT, in which an MBIR reconstruction performed on previous measurements is used to define an objective function on orientations. 
We formally define the orientation selection criterion as a joint optimization problem between (a) a data-fitting term promoting alignment between measurement direction and dominant edges in the reconstruction and (b) a regularization term promoting orientation diversity. 
Using simulated and experimental data, we demonstrate that our adaptive orientation selection method performs better than MBIR with golden ratio sampling \cite{kaestner2011spatiotemporal,venkatakrishnan2021improved} in terms of the convergence speed of the normalized root mean square error (NRMSE), thereby enabling a significant reduction in measurement time for a given reconstruction quality.  

\section{Methodology}

In tomographic imaging, radiograph measurements are sequentially collected from different view angles/orientations during the scanning process. 
In order to design a method for dynamic selection of new orientation, we make two key observations:
\begin{enumerate}
  \item Orientations aligned with the edges of the object of interest contain more information.
  \item An orientation contains less information if a similar orientation has been scanned previously.
\end{enumerate}
The proposed orientation selection method is designed to select new view angles to balance these competing considerations.  

\subsection{Optimization problem formulation}
Let $\theta_n$ be the $n^{th}$ view angle to be selected, $\Theta^*_{n-1}=\left\{\theta^*_0,...,\theta^*_{n-1}\right\}$ be the set of previously selected view angles, and $x_{n}$ be the preliminary reconstruction from CT projections collected at angles $\Theta^*_{n-1}$. 
The angle selection problem is formulated as a joint optimization problem:
\begin{equation}
    \label{optim_problem}
    \theta^*_n=\argmax_{\theta_n \in [0,\pi)}\left\{f(\theta_n;x_{n})+\gamma h(\theta_n;\Theta^*_{n-1})\right\},
\end{equation}
where $f(\theta;x)$ is a data fitting term promoting edge alignment, $h(\theta, \Theta^{*})$ is a regularization term promoting angle diversity, and $\gamma > 0$ is a scalar weight controlling the importance of regularization relative to data fitting.

\subsection{Edge alignment function}
\begin{algorithm}
\caption{Edge Alignment Function Calculation}\label{EA_algorithm}
\textbf{Input}: View angle: $\theta \in [0,\pi)$ \\
\textbf{Input}: Preliminary reconstruction: $x \in \mathbb{R}^{N_z \times N_x \times N_y}$ \\
\textbf{Output}: Edge alignment function: $f(\theta;x) \in [0,1]$ \\
(Note that $x[k] \in \mathbb{R}^{N_x \times N_y}$ is the $k$th slice of $x$.)
\begin{algorithmic}[1]
\State $f(\theta; x)\leftarrow 0$

\For{$k \gets 0$ to $N_z-1$}

\State $\Bar{x}[k] \leftarrow Canny(x[k])$ \algorithmiccomment{Canny edge detection}
\State $\left\{ l_0, ..., l_{M_k-1} \right\}\leftarrow PPHT(\Bar{x}[k])$ \algorithmiccomment{Hough feature}
\For{$m \gets 0$ to $M_k-1$}
\State Calculate $b_m(\theta)$ \algorithmiccomment{Alignment mask from Eq. \ref{mask}}
\State $f(\theta; x)\leftarrow f(\theta; x) + b_m(\theta) \cdot \Bar{x}[k]$
\EndFor
\EndFor
\State $f(\theta;x)\leftarrow \frac{f(\theta; x)}{\max_{\phi \in [0,\pi)}\left\{f(\phi; x)\right\}}$ \algorithmiccomment{Normalization}

\State \Return $f(\theta;x)$
\end{algorithmic}
\end{algorithm}

The edge alignment function $f(\theta; x)$ (Algorithm \ref{EA_algorithm}) determines how well the view angle $\theta$ is aligned with the edges of the attenuation image $x$. 
First, we apply the Canny edge detection algorithm \cite{canny1986computational} to each slice $x[k] \subset x$ to give a binary edge image slice $\bar{x}[k]$, where 1 indicates an edge pixel in the corresponding slice. 
Next, we apply the Progressive Probabilistic Hough Transform (PPHT) \cite{galamhos1999progressive} to $\Bar{x}[k]$ to yield a collection of line segments $L_k=\left\{l_0, ..., l_{M_k-1} \right\}$, each representing a component of a the edge. 
For each line segment $l_m \in L_k$, we evaluate how well it aligns with the view angle $\theta$ by defining a binary mask, $b_m(\theta)$, which is 1 in a conical neighborhood of $l_m$ and 0 elsewhere.   
More precisely,
let $C_\epsilon(l_m)$ be a double-sided cone with vertex at the center of $l_m$, direction-aligned with $l_m$, and with angle of opening $\epsilon$.  Then 
\begin{equation}
\label{mask}
    b_m(\theta)[i,j] = 
    \begin{cases}
    1,& \quad \textrm{if } \textrm{pixel } (i, j) \textrm{ is in } C_\epsilon(l_m) \\
    0,&  \quad \textrm{otherwise.}\\
    \end{cases} \
\end{equation}

The edge alignment function for $x[k]$ is then the inner product between $\Bar{x}[k]$ and the binary masks $b_m(\theta)$, summed over $m=0,...,M_k-1$.  This is then summed over all slices and normalized to give
\begin{equation}
    f(\theta; x) = \frac{1}{K_f(x)} \sum_{k=0}^{N_z-1} \sum_{m=0}^{M_k-1} b_m(\theta)\cdot\Bar{x}[k], 
\end{equation}
where $K_f(x)$ is chosen so that $\max_{\theta \in [0,\pi)} f(\theta;x)=1$.

\begin{figure}[t]%
     \centering
     \begin{subfigure}[t]{0.39\columnwidth}
         \centering
         \includegraphics[width=\textwidth]{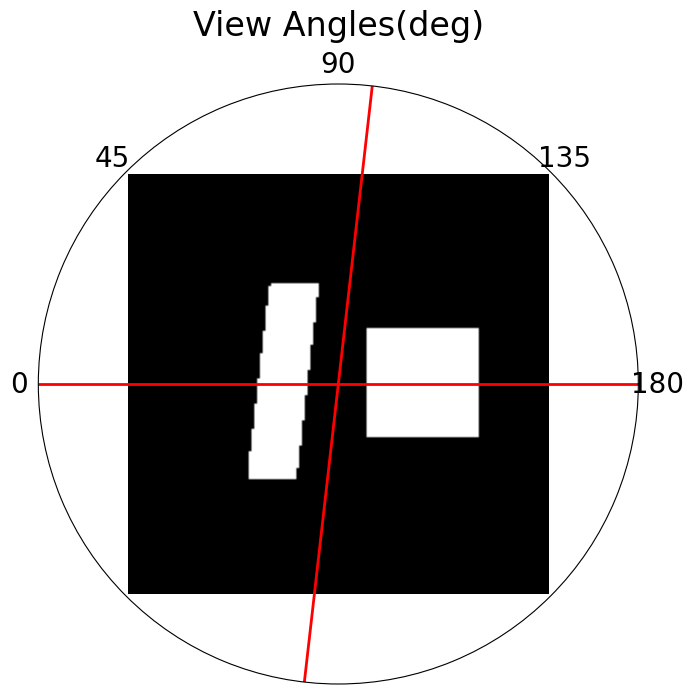}
         \caption{Single image slice}
         \label{fig: EA_attenuation}
     \end{subfigure}
     \begin{subfigure}[t]{0.59\columnwidth}
         \centering
         \includegraphics[width=\textwidth]{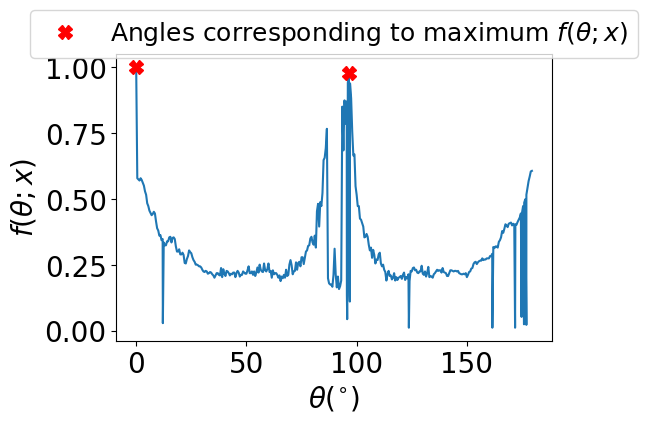}
         \caption{Plot of edge alignment function}
         \label{fig: EA_function_plot}
     \end{subfigure}
    \caption{(a) Illustration of an binary image of size 150$\times$150 pixels to demonstrate the edge alignment terms in the cost function. 
    Red lines indicate the angles at which the edge alignment function is maximized. 
    (b) Edge alignment function vs.~view angle, showing angles for which the function is maximized (i.e. best aligned with existing edges).
    Notice that the function is maximum at the orientations that are aligned with the dominant edges of the image. 
    }
    \label{fig: EA_visualization}
\end{figure}

Fig. \ref{fig: EA_attenuation} shows an example of a binary image, and Fig. \ref{fig: EA_function_plot} shows the corresponding plot of the edge alignment function. In this example $f(\theta;x)$ takes on maximum values at $\theta=0^\circ$ and $95^\circ$, with a slightly smaller peak at $90^\circ$, corresponding to the orientations of the dominant edges in the attenuation slice.

\subsection{Angle spacing cost function}
Given a view angle $\theta$ and all previously selected view angles $\Theta^*=\left\{\theta^*_0,...,\theta^*_{n-1}\right\}$, the angle spacing function is defined as
\begin{equation}
    \label{angle_spacing_function}
    h(\theta;\Theta^*)= \frac{1}{K_h(\Theta^*)} \exp(-\alpha \sum_{j=0}^{n-1}\frac{1}{d(\theta,\theta^*_j)}),
\end{equation}
where $K_h(\Theta^*)$ is chosen so that $\max_\theta h(\theta;\Theta^*) = 1$, 
$d(\theta,\theta^*_j)=\min\left\{\abs{\theta-\theta^*_j}, 180^\circ-\abs{\theta-\theta^*_j}\right\}$ is  angular distance, and $\alpha$ is a scaling parameter to control the importance of proximity to previously measured angles.  

\begin{figure}[t]
\centering
   \includegraphics[width=0.7\linewidth]{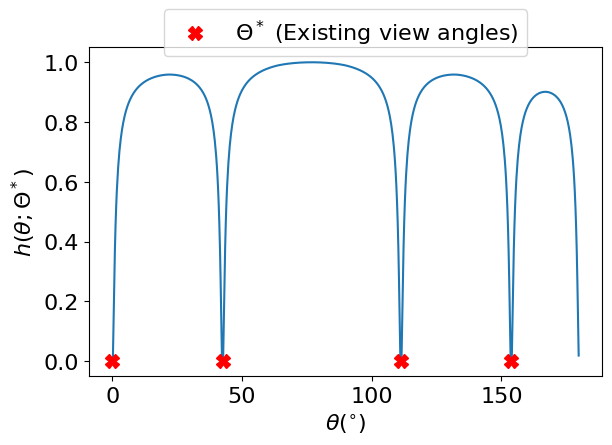}
   \caption{Normalized angle spacing cost function vs. view angles showing existing view angles (red marks).
   We note that the cost function strongly discourages the selection of angles that are close to previously selected angles.}
   \label{fig:AS_example}
\end{figure}

Fig.~\ref{fig:AS_example} shows an example plot of $h(\theta;\Theta^*)$ with four existing view angles. 
As $\theta$ approaches an existing view angle, $h$ decreases quickly, while $h$ is relatively constant for $\theta$ values reasonably spaced from all existing angles.

\subsection{Parameter selection}
One interpretation of the joint optimization problem in Eq. \ref{optim_problem} is as an exploration-exploitation trade-off. 
Edge alignment, promoted by $f(\theta;x)$, encourages exploitation of existing edge information, while angle spacing, promoted by $h(\theta;\Theta^*)$, encourages exploration of new orientations, and the relative importance of each is governed by $\gamma$. 
While the choice of $\gamma$ could be influenced by prior knowledge of the prevalence of edges in a sample or by preliminary reconstructions, we use $\gamma=1$ for simplicity and to establish a baseline for future work.
Likewise, we choose $\alpha=1$ in Eq.~\ref{angle_spacing_function} for simplicity and to establish a baseline for future work.

\section{Numerical Implementation}

We solve Eq. \ref{optim_problem} numerically using a discretized solution space $\Theta=\left\{0, \Delta\theta, ..., \pi-\Delta\theta\right\}$.
Algorithm \ref{EAVS_algorithm} shows the pseudocode of the proposed orientation selection algorithm, which first calculates a reconstruction $x_n$ using previous angles, then evaluates the functions $f$ and $h$ over the discrete set $\Theta$, then does the optimization.
We obtain the reconstruction $x_n$ from $\Theta^*_{n-1}$ by using the previous reconstruction $x_{n-1}$ as initialization for the svMBIR reconstruction package \cite{svmbir}, which rapidly yields high-quality reconstructions with an automated choice of regularization parameters. 
As a result, the angle selection of Algorithm~\ref{EAVS_algorithm} completes quickly relative to the long measurement time of each projection measurement in nCT.
Algorithm \ref{simulation_algorithm} describes the complete angle selection and reconstruction workflow, in which  an initial reconstruction is performed, a view angle is selected, the corresponding projection is measured, and the reconstruction is performed with all measured projections.
\begin{algorithm}
\caption{Single-Step Orientation Selection Algorithm}\label{EAVS_algorithm}
\textbf{Input}: Previously measured angles: $\Theta^*_{n-1}$ \\
\textbf{Input}: Previous reconstruction: $x_{n-1}$ \\
\textbf{Input}: Discretized candidate angles: $\Theta_n$ \\
\textbf{Output}: Next view angle: $\theta^*_{n}$ 
\begin{algorithmic}[1]
\State Calculate $f(\theta;x_{n-1})$, $\forall \theta \in \Theta_n$ using  Algorithm \ref{EA_algorithm}
\State Calculate $h(\theta;\Theta^*_{n-1})$, $\forall \theta \in \Theta_n$ using Eq.~\ref{angle_spacing_function}
\State $\theta^*_{n} \leftarrow \argmax_{\theta \in \Theta_n}\left\{ f(\theta;x_{n-1})+\gamma h(\theta;\Theta^*_{n-1})\right\}$
\State \Return $\theta^*_n$
\end{algorithmic}
\end{algorithm}

\begin{algorithm}
\caption{Proposed Neutron Imaging Workflow} \label{simulation_algorithm}
\textbf{Input}: Initial view angles: $\Theta^*_{0}$ \\  
\textbf{Input}: Discretized view angle space: $\Theta$ \\  
\textbf{Input}: Initial measurement data: $Y_{0}$ \\ 
\textbf{Input}: Number of additional views to acquire: $N$ \\
\textbf{Output}: Final Reconstruction: $x_N$ 
\begin{algorithmic}[1]
\State $x_{0}\leftarrow \texttt{Recon}(Y_{0}, \Theta^*_{0})$ \algorithmiccomment{Initial reconstruction}
\For{$n \gets 1$ to $N$}
    \State $\Theta_n \gets \Theta \setminus \Theta^*_{n-1}$  \algorithmiccomment{Update candidate view angles}
    \State $\theta^*_{n}\leftarrow \texttt{SelectAngle}(x_{n-1}, \Theta^*_{n-1}, \Theta_n)$ \algorithmiccomment{Algo.~\ref{EAVS_algorithm}}
    \State $y_{n} \leftarrow $ measurement of  projection at angle $\theta^*_{n}$
    \State $Y_{n} \leftarrow \{Y_{n-1}, y_{n}\}$ \algorithmiccomment{Update measurement data}
    \State $\Theta^*_{n}\leftarrow \{\Theta^*_{n-1}, \theta^*_{n}\}$ \algorithmiccomment{Update existing view angles}
    \State $x_{n} \leftarrow \texttt{Recon}(Y_{n}, \Theta^*_{n})$ \algorithmiccomment{Update reconstruction}
\EndFor
\State \Return $x_{N}$
\end{algorithmic}
\end{algorithm}

\section{Results}
\iftrue
In this section, we present experimental results from both synthetic and real nCT scans of various 3D objects.  In each case, we compare reconstructions using the proposed orientation selection algorithm to reconstructions using MBIR in conjunction with the golden ratio acquisition method.  

For synthetic 3D data, we generated two binary ground truth volumes composed of multiple geometric shapes such as sphere, prism, cube, etc. 
Each ground truth volume has 50 axial slices, with 150$\times$150 pixels per slice. 
We then scaled the ground truth volume by a factor of $0.01$ in order to make the attenuation values physically realistic. 
For each ground truth volume, we generated a synthetic data by first forward projecting the volume at selected view angles using svMBIR's \cite{svmbir} parallel beam projector and then modeling the measurement using Poisson counting with values similar to typically observed real experiments. 

The experimental data is from a sample of a volcanic rock held on a metal ring, with densely sampled measurements obtained at the imaging instrument High Flux Isotope Reactor (HFIR) at Oak Ridge National Lab (ORNL). 
For the NRMSE calculation, we used an MBIR reconstruction from 1200 projections to generate a reference volume with 5 axial slices and 400$\times$400 pixels per slice.

We compare the proposed algorithm to the golden ratio method in Fig.~\ref{fig:simulated_results_display} (simulated data) and Fig.~\ref{fig:real_results_display} (real data). 
In both cases, the proposed algorithm outperforms the golden ratio method in terms of the convergence speed of NRMSE between the reconstruction and the reference volume.  
This improvement is especially pronounced in the early stages of angle selection in  Fig.~\ref{fig:simulated_results_display}(e) and Fig.~\ref{fig:real_results_display}, in which there are prominent edges that are better resolved by appropriate angle selection, which in turn yields a significantly better reconstruction.  In the case of Fig.~\ref{fig:simulated_results_display}(f), the object has fewer clearly defined edges, and the initial reconstruction from 3 view angles is better than that in  Fig.~\ref{fig:simulated_results_display}(e), so the improvement over the golden ratio method is still apparent, but less pronounced.  

\begin{figure}[t]%
     \begin{subfigure}[t]{0.49\columnwidth}
         \centering
         \includegraphics[width=\textwidth]{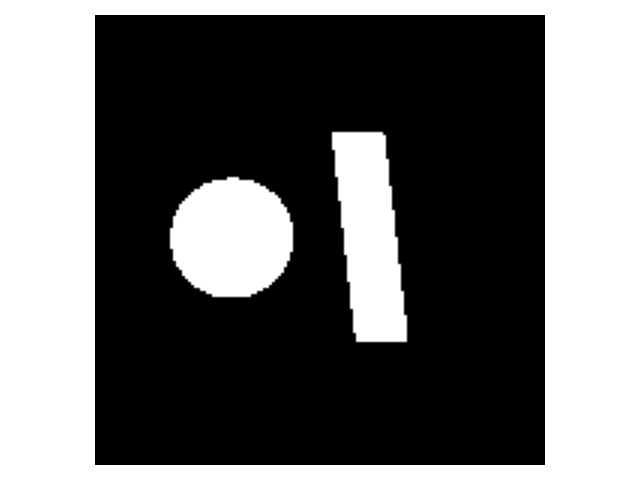}
         \label{fig: phantom_mod1}
         \vskip -5mm
         \caption{Ground truth}
     \end{subfigure}
     \begin{subfigure}[t]{0.49\columnwidth}
         \centering
         \includegraphics[width=\textwidth]{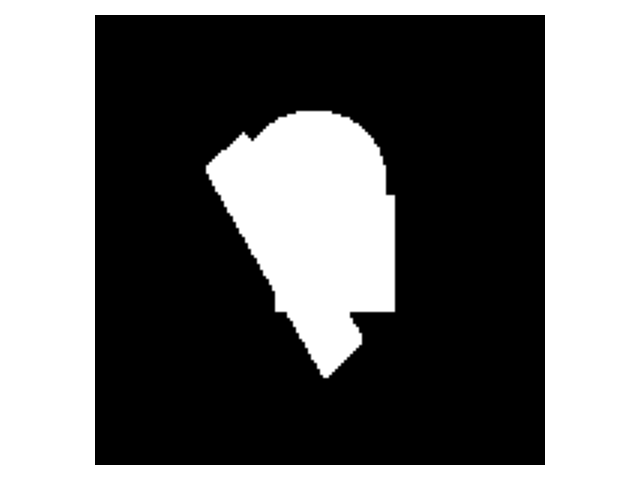}
         \label{fig: phantom_mod2}
         \vskip -5mm
         \caption{Ground truth}
     \end{subfigure}
     \\
     \vskip 5mm
     \begin{subfigure}[t]{0.49\columnwidth}
         \centering
         \includegraphics[width=\textwidth]{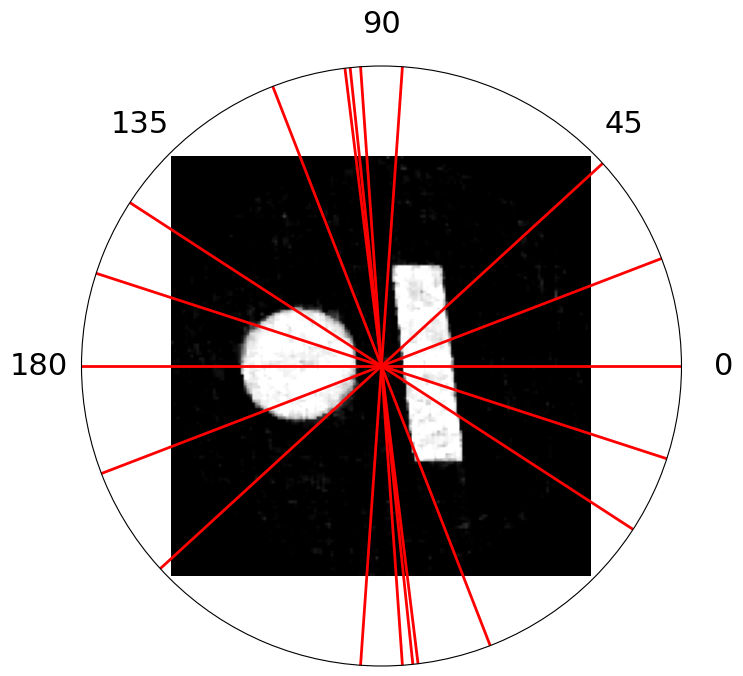}
         \label{fig: recon_mod_1_noisy}
         \vskip -5mm
         \caption{Selected angles and reconstruction}
     \end{subfigure}
      \begin{subfigure}[t]{0.49\columnwidth}
         \centering
         \includegraphics[width=\textwidth]{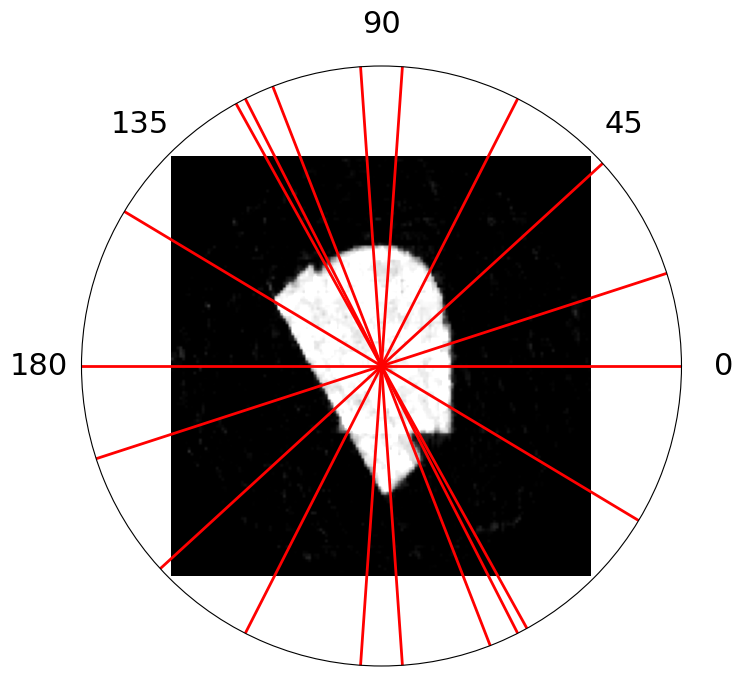}
         \label{fig: recon_mod_2_noisy}
         \vskip -5mm
         \caption{Selected angles and reconstruction}
     \end{subfigure}
     \\
     \vskip 5mm
     \begin{subfigure}[b]{0.49\columnwidth}
         \centering
         \includegraphics[width=\textwidth]{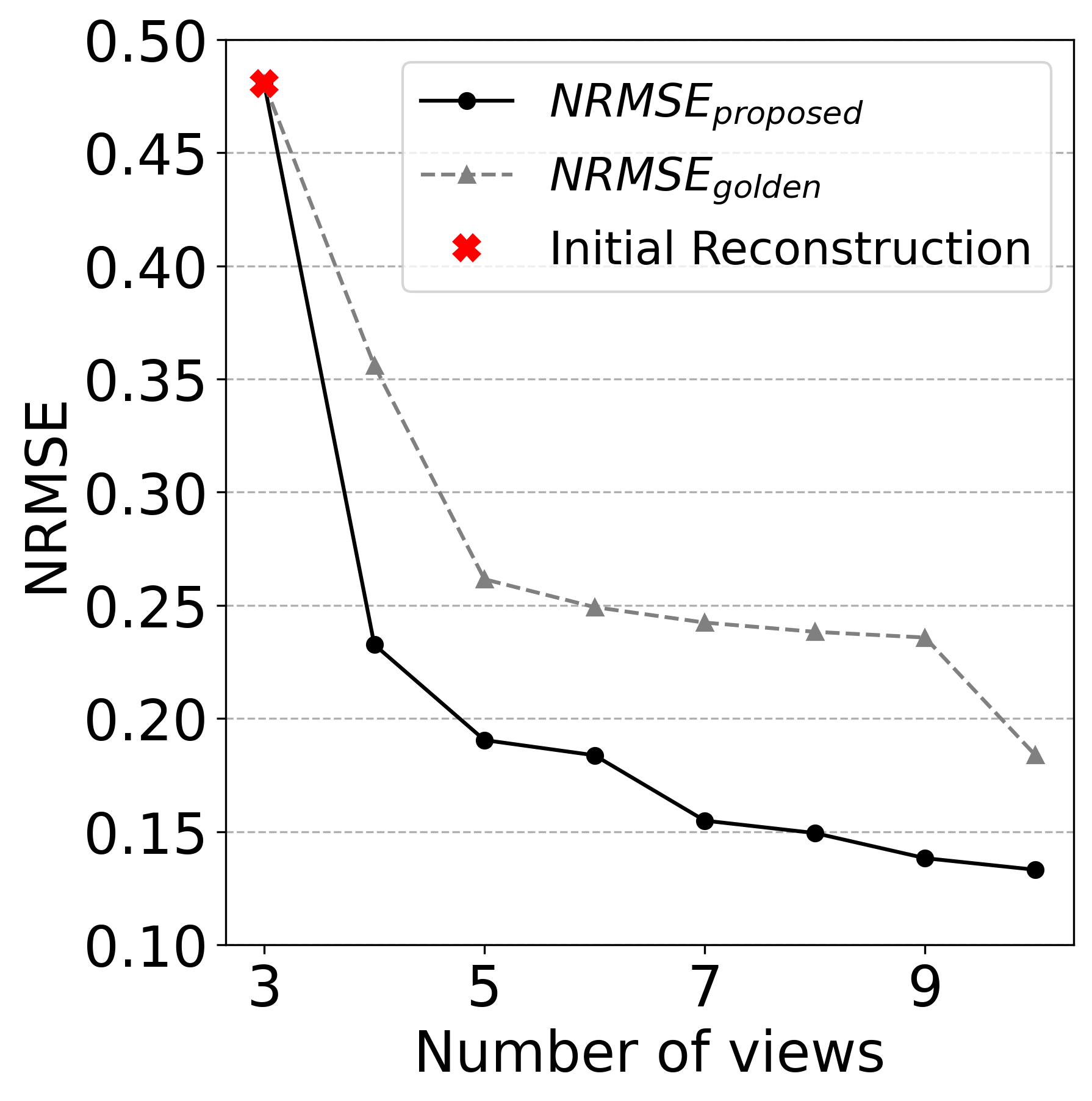}
         \vskip -1mm
         \caption{NRMSE vs \#views}
         \label{fig: nrmse_mod_1_noisy}
     \end{subfigure}
     \begin{subfigure}[b]{0.49\columnwidth}
         \centering
         \includegraphics[width=\textwidth]{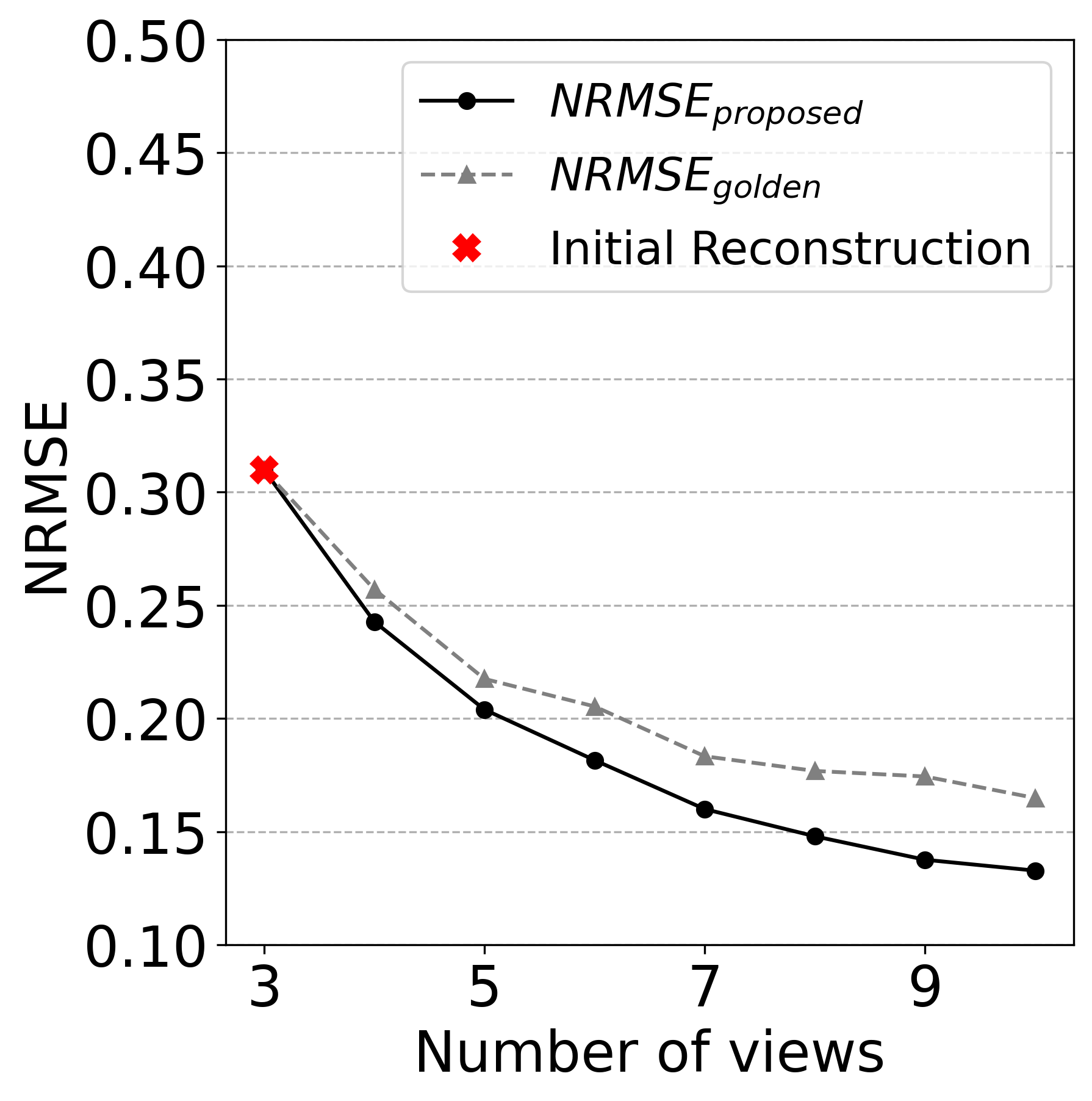}
         \vskip -1mm
         \caption{NRMSE vs \#views}
         \label{fig: nrmse_mod_2_noisy}
     \end{subfigure}
        \caption{Results using simulated data. (a) and (b) show representative slices from ground truth volumes;
        (c) and (d) show the angles selected by the proposed algorithm and the corresponding reconstructions using MBIR; (e) and (f) show the corresponding NRMSE as a function of the number of view angles for each of the golden ratio method and the proposed method. 
        For both sample geometries, the NRMSE converges faster when using the proposed method.}
        \label{fig:simulated_results_display}
\end{figure}


\begin{figure}[t]%
     \begin{subfigure}[t]{0.49\columnwidth}
         \centering
         \includegraphics[width=\textwidth]{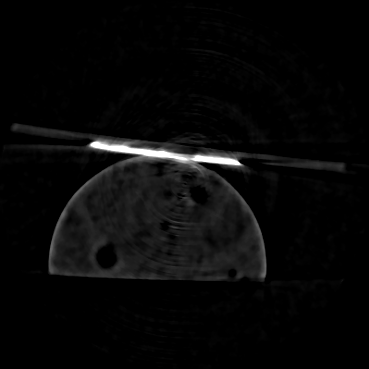}
         \caption{Reference image}
         \label{fig: real_data_ground_truth}
     \end{subfigure}
     \begin{subfigure}[t]{0.49\columnwidth}
         \centering
         \includegraphics[width=\textwidth]{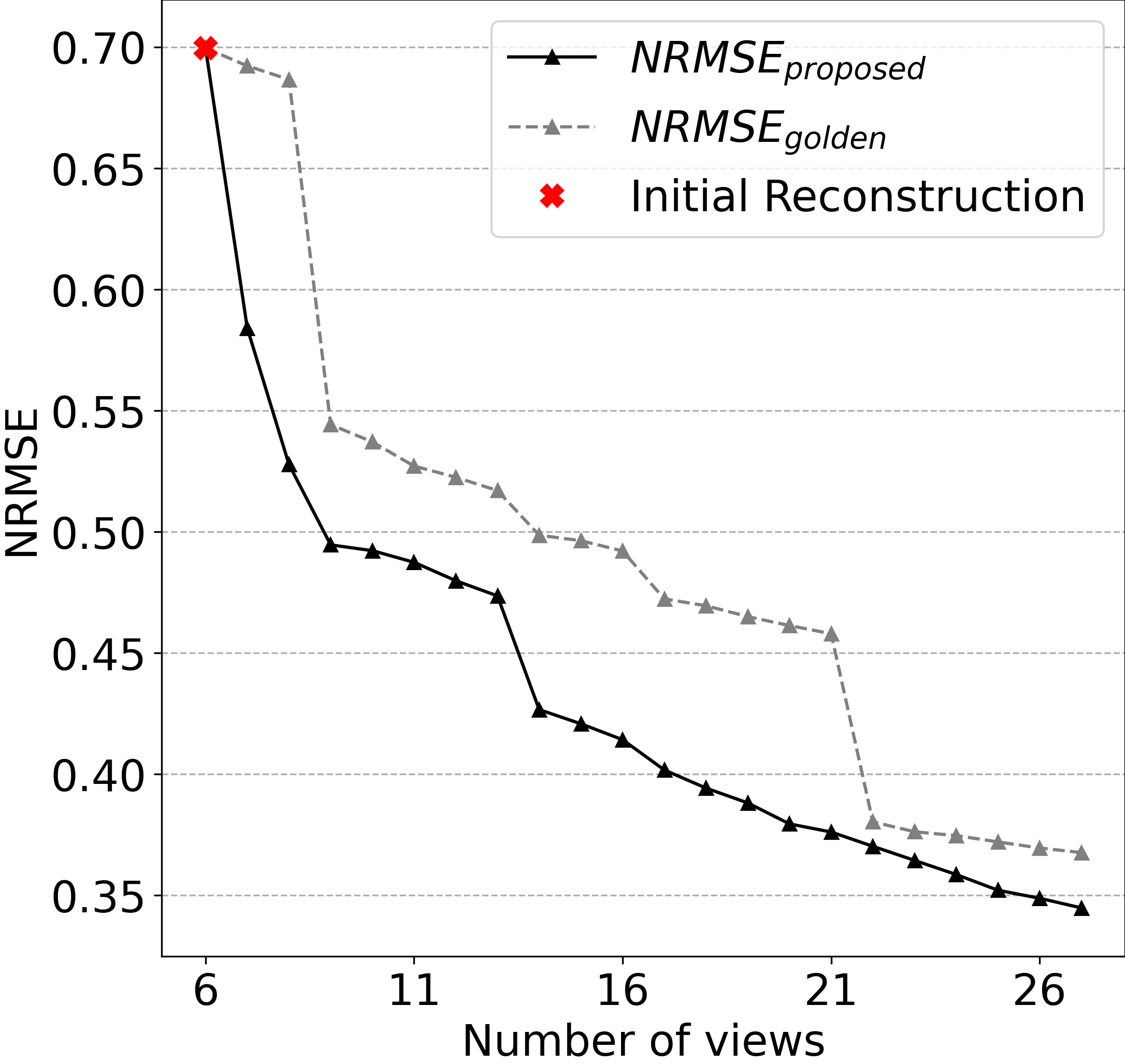}
         \caption{NRMSE vs \#views}
         \label{fig: real_data_nrmse}
     \end{subfigure}
     \\
     \vskip 5mm
     \begin{subfigure}[t]{0.49\columnwidth}
         \centering
         \includegraphics[width=\textwidth]{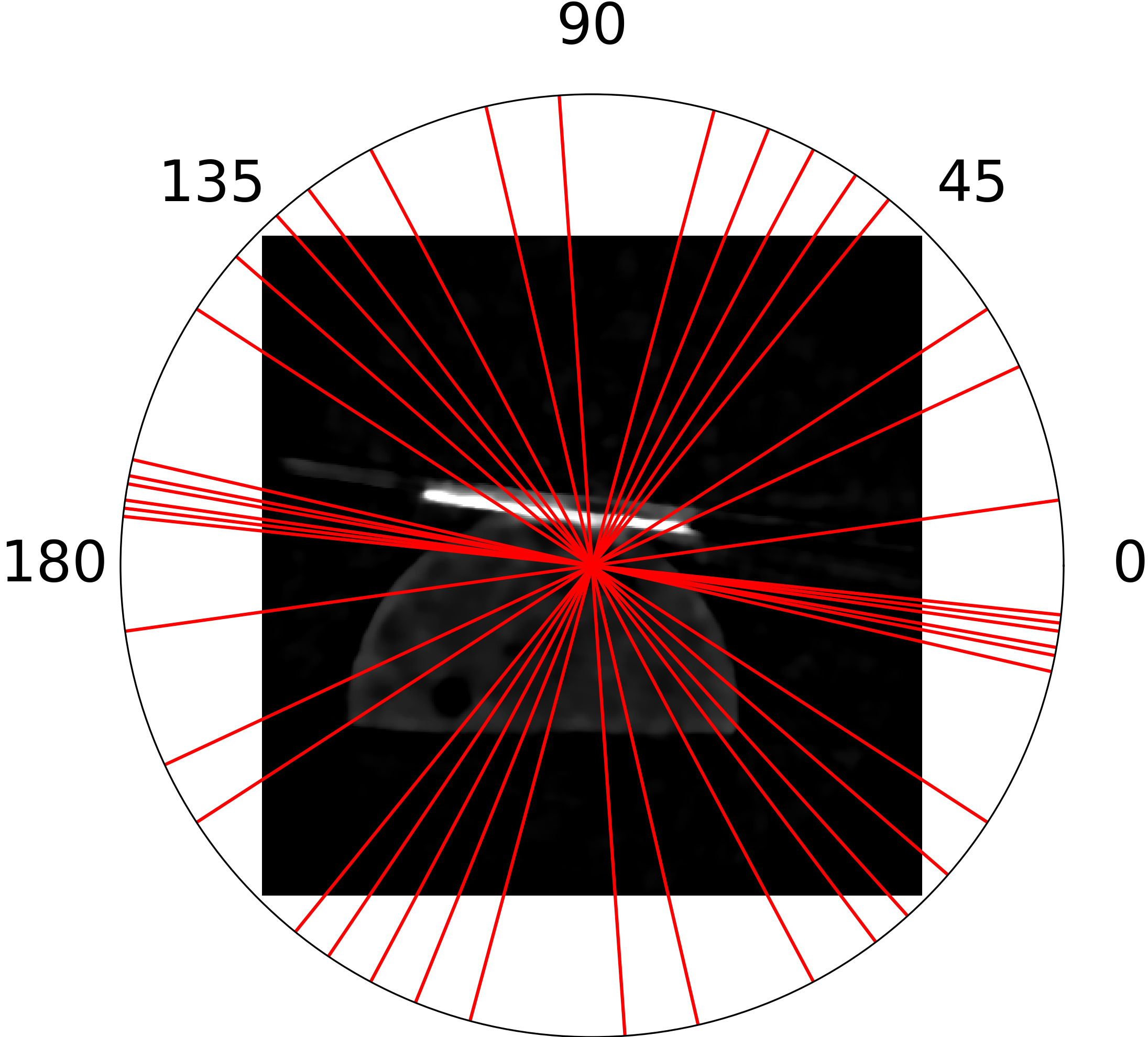}
         \caption{Proposed method: Selected angles and reconstruction with 20 projections}
         \label{fig: real_data_eavs_res}
     \end{subfigure}
     \begin{subfigure}[t]{0.49\columnwidth}
         \centering
         \includegraphics[width=\textwidth]{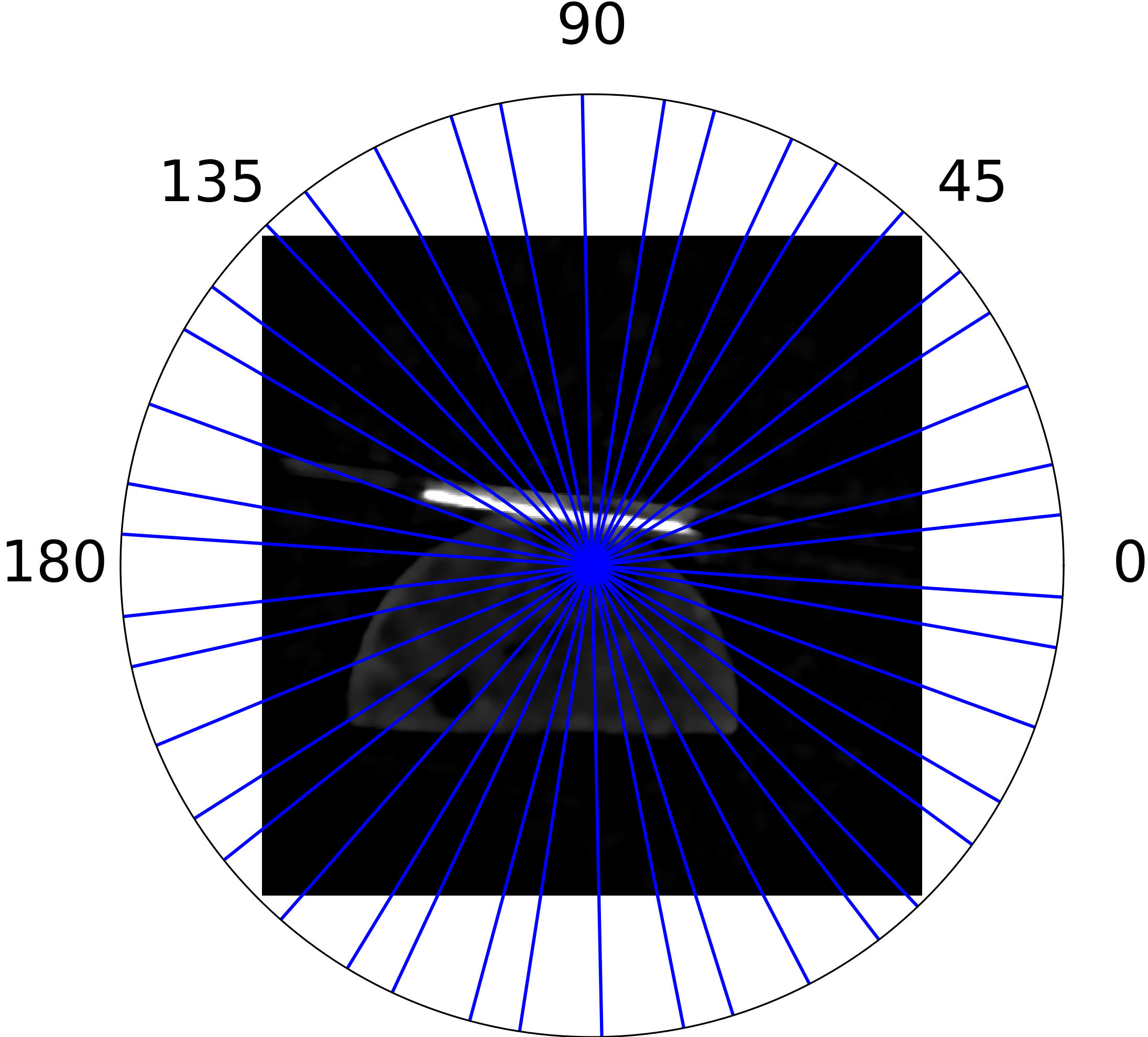}
         \caption{Golden ratio: Selected angles and reconstruction with 20 projections}
         \label{fig: real_data_golden_res}
     \end{subfigure}
        \caption{Results using experimental data.
        (a) shows the reference image (MBIR from 1200 projections).
        (b) shows the NRMSE as a function of number of view angles for the proposed method and the golden ratio method. 
        (c) and (d) respectively show the first 20 angles selected by the proposed method and the golden ratio method (after 6 evenly spaced initial angles) and the corresponding reconstructions using MBIR.
        We note that the proposed orientation selection method leads to a faster convergence to the reference compared to the traditional golden ratio method.
        } 
        \label{fig:real_results_display}
\end{figure}

\section{Conclusion}
In this paper, we introduced a sample adaptive measurement strategy that can significantly reduce the measurement time of neutron CT scans. 
Our method uses existing measurements to perform an MBIR reconstruction, which is then used to construct an edge image to locate prominent edges in the object.  The reconstruction and edge image are used to define the orientation selection criterion as a joint optimization problem between a data-fitting term promoting edge alignment and a regularization term promoting angle spacing to ensure a diversity of views.
Using simulated and experimental data, we demonstrated that our orientation selection method performs better than MBIR applied with the conventional golden ratio sparse acquisition scheme in terms of the convergence speed of NRMSE, thereby yielding a significant reduction in measurement time for a given quality of reconstruction.


\section*{Acknowledgment}
A portion of this research used resources at the High Flux Isotope Reactor and Spallation Neutron Source, DOE Office of Science User Facilities operated by the Oak Ridge National Laboratory. Oak Ridge National Laboratory is managed by UT-Battelle, LLC, for the U.S. DOE under contract DE-AC05-00OR22725. 
We give special thanks to Professor Mattia Pistone from University of Georgia, who provided permission for use of the volcanic rock dataset.
\newpage
\bibliographystyle{IEEEtran}
\bibliography{icassp.bib}

\end{document}